# Inferential Validity of Digital Health Measures

Carmen D. Tekwe[1*], Mercy Oladuti[1], Yuanyuan Luan[2], Andi Mai[1], Orfeu M. Buxton[3], See Ling Loy[4,5], Jeffrey S. Gonzalez[6,7,8,9,10], Roger S. Zoh[1*]

[1]Department of Epidemiology and Biostatistics, Indiana University School of Public Health-Bloomington, Bloomington, IN, USA

[2]Washington University in St. Louis Bursky School of Public Health, Saint Louis, MO, USA

[3]Department of Biobehavioral Health, College of Health and Human Development, Pennsylvania State University, University Park, PA, USA

[4]Department of Reproductive Medicine, KK Women's and Children's Hospital, Singapore, Singapore

[5]Obstetrics & Gynecology Academic Clinical Program, Duke-NUS Medical School, Singapore, Singapore

[6]Ferkauf Graduate School of Psychology, Yeshiva University, Bronx, NY, USA

[7]Department of Medicine (Endocrinology), Albert Einstein College of Medicine, Bronx, NY, USA

[8]Department of Epidemiology and Population Health, Albert Einstein College of Medicine, Bronx, NY, USA

[9]Fleischer Institute for Diabetes and Metabolism, Department of Medicine, Albert Einstein College of Medicine, Bronx, NY, USA

[10]New York-Regional Center for Diabetes Translation Research, Albert Einstein College of Medicine, Bronx, NY, USA

* Correspondence: Carmen D. Tekwe, ctekwe@iu.edu ; Roger S. Zoh, rszoh@iu.edu

ORCID iDs

Carmen D. Tekwe: 0000-0002-1857-2416

Mercy Oladuti: 0000-0001-9852-2066

Yuanyuan Luan: 0000-0003-1861-5292

Andi Mai: 0009-0006-5889-0490

Orfeu M. Buxton: 0000-0001-5057-633X

See Ling Loy: 0000-0002-9205-5580

Jeffrey S. Gonzalez: 0000-0002-8252-2077

Roger S. Zoh: 0000-0002-8066-1153

**Abstract**

Digital health measures increasingly inform treatment evaluation, risk classification, clinical monitoring, and regulatory decisions. Existing validity concepts concern a measure's technical soundness, clinical meaningfulness, usability, and scalability. Inferential validity concerns how well a measure supports a scientific, clinical, or regulatory conclusion. We propose assessing inferential validity as a downstream evaluation layer that links measurement system evidence to trustworthy conclusions.

## Introduction

Digital health technologies are rapidly changing how patients and professionals measure and monitor health and act in response to the results. Digital health measures now often inform treatment decisions, risk classification, assessments of therapeutic effectiveness, and regulatory decisions. As these measures increasingly underlie clinical and regulatory decisions, confidence in the conclusions they support becomes as important as confidence in the measurements themselves. Continuous glucose monitors, physical activity monitors, optical heart rate sensors, sleep wearables and devices, and remote monitoring platforms now generate dense measures of physiology and behavior in daily life. Researchers and clinicians use these measures to study disease progression, evaluate interventions, develop digital biomarkers, monitor patients remotely, and trigger clinical and behavioral decisions[1-3]. Their value is clear: they allow observation outside clinic walls in more ecologically valid contexts, and capture patterns that occasional visits or self-reports may miss.

A digital measure can appear accurate, stable, or predictive yet still provide weak support for a particular conclusion. Continuous glucose monitors (CGMs) may perform well on average but disagree on whether measurements are above or below thresholds that guide diabetes management[4,5]. A physical activity monitor may provide repeatable daily summaries but undercount some movement patterns, weakening estimated links between activity and cardiometabolic health[6,7]. A pulse oximeter may agree with arterial oxygen saturation on average while missing low oxygen events more often in some groups than others[8]. These examples illustrate that confidence in a digital measure does not necessarily translate into confidence in the conclusions drawn from it[9].

Existing validity concepts provide essential structure. Verification refers to whether a device captures the intended raw signal. Analytical validation refers to how accurately a measure represents a target quantity under specified conditions, such as a defined device version, wear location, population, and setting. Clinical validation refers to how meaningfully related a measure is to a clinical state, outcome, or construct[10,11]. The V3+ framework builds on these concepts by adding user-centered design and scalability, because digital health technologies must be usable by intended users and deployable in real-world settings[1,10].

Although these concepts are essential, they do not address the pathway from digital measure to conclusion fully. This gap motivates inferential validity, which we define as the extent to which a digital measure supports a trustworthy scientific, clinical, or regulatory conclusion after accounting for the measurement system that generated the measure and the context of the conclusion. The measurement system includes the device, sensor hardware, firmware, software, preprocessing choices, calibration, body placement, user behavior, missing data, environmental conditions, and changes over time[12-16]. Inferential validity can be evaluated by considering three related questions: how well the digital measure corresponds to the scientific, clinical, or regulatory question; how measurement system uncertainty could change the conclusion; and how stable the conclusion remains across devices, algorithms, populations, workflows, software versions, and deployment settings. Alignment refers to the correspondence between the intended question and the digital measure used to answer it. Systematic sensor differences are a source of measurement system uncertainty that can alter measured values and change downstream classifications or conclusions when threshold-based decision rules are applied.

Figure 1 illustrates this inferential pathway, beginning with the measurement system and proceeding through the digital measure and analysis or decision rule to the resulting scientific, clinical, or regulatory conclusion. Inferential validity concerns the strength of support for the conclusion across this pathway. Inferential failure can arise when device characteristics, algorithm or firmware changes, preprocessing choices, missing data, modeling assumptions, subgroup differences, or deployment contexts alter the conclusion.

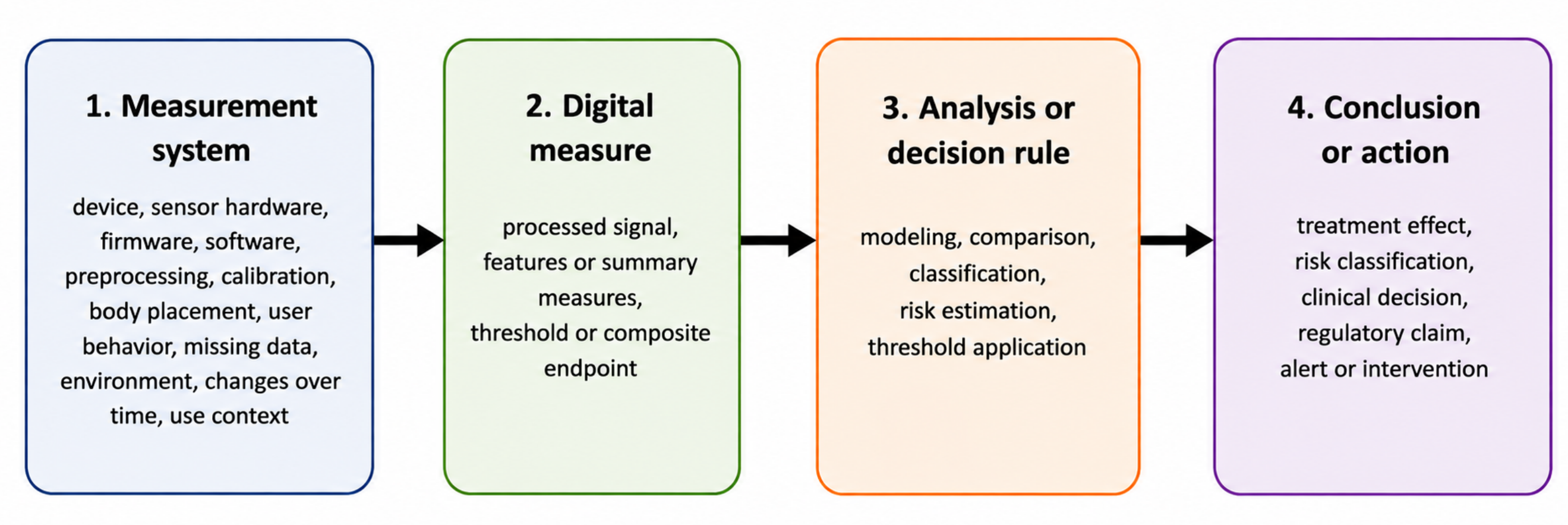

**Figure 1. Inferential validity across the digital health measurement pipeline.** The pathway links the measurement system, digital measure, analysis or decision rule, and conclusion or action. Inferential validity concerns the strength of support for the conclusion throughout this pathway. Evaluation of inferential validity considers (1) whether the digital measure is aligned with the intended scientific, clinical, or regulatory question, (2) whether measurement system uncertainty could alter the conclusion, and (3) whether the conclusion remains stable across devices, algorithms, populations, workflows, software versions, and deployment settings.

## Illustrative examples

We illustrate inferential validity with three common uses of digital measures: estimating an association, classifying a person relative to a threshold, and comparing performance across subgroups. These examples show how measurement system behavior can change the conclusion even when a measure appears useful overall.

### *Physical activity and adiposity*

Researchers have used accelerometry-derived physical activity trajectories from the National Health and Nutrition Examination Survey (NHANES) to estimate how activity across the day relates to adiposity[17-23]. Different approaches to handling imperfect wearable measurements have produced different conclusions about when activity is most strongly associated with adiposity. Methods that reconstruct smoother activity trajectories from noisy observed wearable data[24,25] can produce different 24-hour association curves than methods that correct measurement error explicitly[26].

This example highlights a simple point: smoothing a noisy activity curve differs from correcting a biased measure. Smoothing can make a trajectory look more coherent while preserving systematic error from device limitations, wear location, or processing choices. In simulation studies, correcting for measurement error reduced mean integrated squared error by approximately 80% to 90% relative to an uncorrected estimator, indicating better recovery of the timing and magnitude of the activity-adiposity association[26]. This example shows how different approaches to handling measurement system uncertainty can lead to different estimates of the activity-adiposity relationship and different scientific conclusions.

### *CGM and clinical threshold decisions in diabetes*

CGM is a standard of care for most people with type 1 diabetes and is becoming common for individuals with type 2 diabetes, particularly those receiving insulin or other therapies that can cause hypoglycemia[43]. Clinical practice and patient self-management rely on threshold-based CGM summaries, including time in range and time below range, to evaluate treatment safety and effectiveness[4,43,44]. A common threshold for low glucose exposure is 70 mg/dL, and more than 4% of monitored time below this value indicates clinically important hypoglycemia burden[4,44]. Hypoglycemia can have serious consequences and is a particular concern in older or frail adults, who may be more vulnerable to falls, cognitive effects, cardiovascular events, and treatment burden[44,45]. Clinicians therefore use time below range to guide treatment review and compare therapies in terms of their risk in exposing patients to low glucose levels.

Using the CGMacros paired device CGM dataset[27,28], we examined how disagreement between simultaneously worn Abbott FreeStyle Libre Pro and Dexcom G6 Pro sensors can change the interpretation of time below range near 70 mg/dL. In the participant-day shown in Figure 2A, Libre values stayed below 70 mg/dL for almost the entire day, while Dexcom values were mostly above 70 mg/dL. That is, during the same monitoring period, one sensor suggested high hypoglycemia burden and the other sensor suggested low burden.

In the CGMacros data, 17 of 45 participants exceeded 4% time below range with Libre measurements, but none of the 45 participants did so with Dexcom measurements. The largest differences in percentage of time below range exceeded 80 percentage points. Measured mean glucose may differ only modestly between devices, but a threshold summary is sensitive because its calculation involves classifying each time point as above or below a boundary. Thus, systematic sensor disagreement near a clinically meaningful threshold could alter classifications, estimates of treatment benefit, eligibility for intervention, or interpretation of hypoglycemia burden. Similar changes in inference could arise from differences in device generations, calibration procedures, time since sensor insertion, sensor placement, firmware or software updates, preprocessing procedures, or other components of the measurement system.

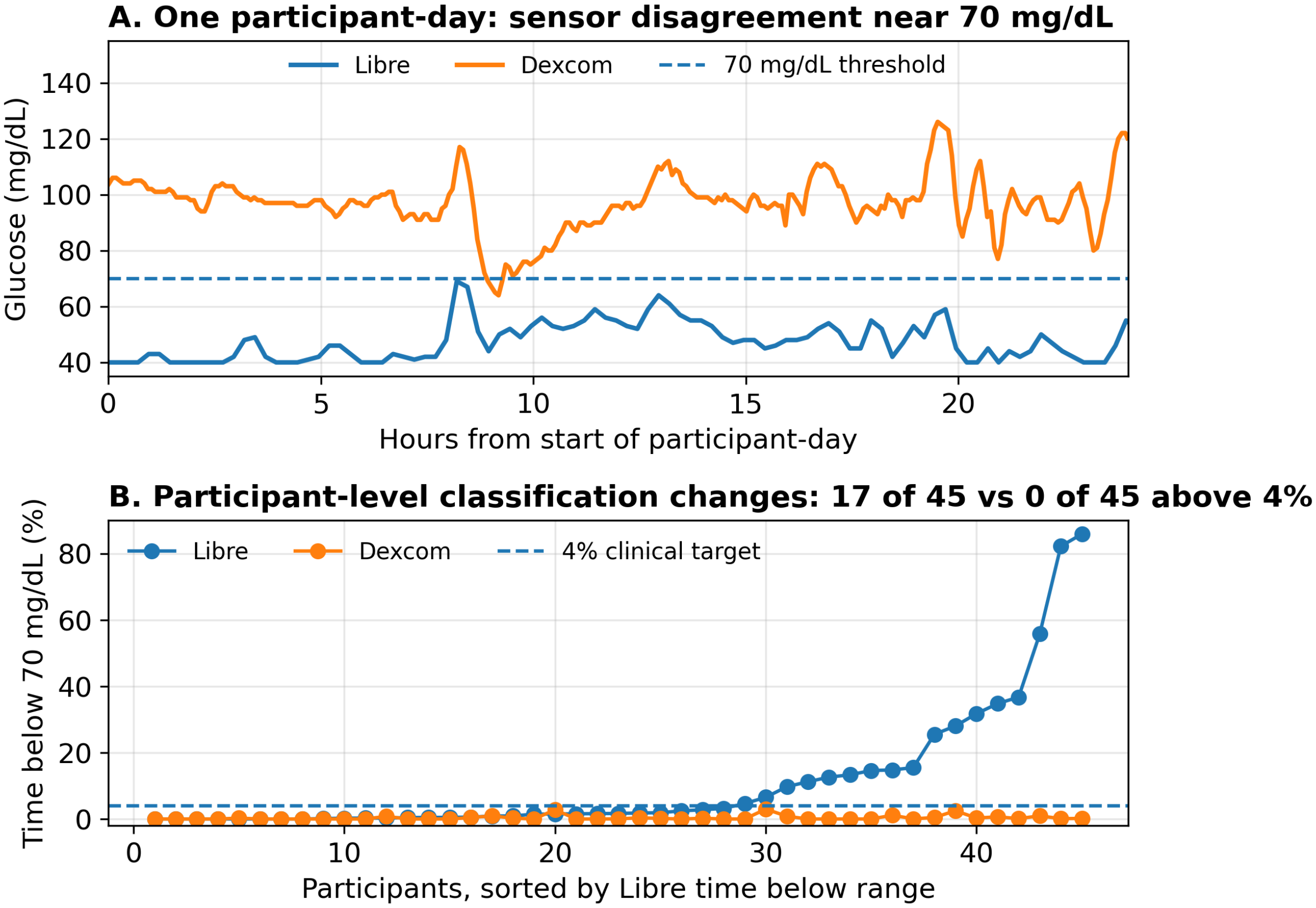

**Figure 2. Near-threshold sensor disagreement can alter interpretation of time below range**. Panel A shows simultaneous Libre and Dexcom glucose measurements for one participant-day, with the 70 mg/dL threshold marked. Panel B shows participant-level time below 70 mg/dL for 45 participants, sorted by Libre time below range.

### *Interpretation of pulse oximetry across subgroups*

Treatment decisions based on pulse oximetry often depend on fixed oxygen saturation thresholds. Occult hypoxemia occurs when pulse oximetry appears reassuring even though arterial oxygen saturation is low. In an illustrative analysis of OpenOximetry data[29-31], pulse oximeter values showed acceptable agreement with arterial oxygen saturation on average, but the direction and size of error differed across skin tone groups.

The researchers defined occult hypoxemia as pulse oximeter values at or above 90% despite arterial oxygen saturation below 90%. Occult hypoxemia occurred in 6.6% of observations in participants with light skin tone, 6.2% in those with medium skin tone, and 18.9% in those with dark skin tone. The dark skin tone group had approximately threefold higher odds of occult hypoxemia than the light skin tone group in adjusted analyses; results were similar for 88% and 92% pulse oximeter thresholds. Thus, measurement system behavior can differ across clinically relevant subgroups, leading to different conclusions about who needs intervention even when overall measurement performance appears acceptable[8,32].

## Common sources of inferential failure across the digital measure pipeline

Problems can emerge throughout the digital measurement pipeline. Missing data, software or firmware updates, algorithm changes, multimodal data fusion, and deployment in a new setting can all change the strength of support for a conclusion. Table 1 maps common sources of inferential failure to the pathway in Figure 1.

**Table 1.** Common sources of inferential failure across the digital measurement pipeline and strategies for strengthening inferential validity.

| Pipeline stage | Source of inferential failure | Concrete example | Effect on inferential validity | Evaluation or mitigation strategy |
|---|---|---|---|---|
| Measurement system | Sensor disagreement, calibration differences, subgroup-specific error, sensor wear duration, or wear conditions | Two CGM systems disagree near 70 mg/dL; pulse oximetry overestimates oxygen saturation in a clinically relevant subgroup | The measured signal can alter classification, treatment effect estimates, or clinical interpretation near a decision boundary | Compare measurements with reference or paired device data; evaluate performance near clinical thresholds and across subgroups; assess calibration, sensor wear duration, and wear conditions |
| Digital measure construction | Preprocessing, calibration, wear location, software version, or summary definition | A software update changes the calculation of a daily activity, glucose, or sleep summary | The derived measure may no longer represent the same target construct or support the interpretation established under an earlier version | Record device, firmware, software, and processing versions; repeat validation after material changes; conduct sensitivity analyses with alternative processing choices |
| Analysis or decision rule | Missingness, adherence, sampling frequency, or model assumptions | Wear time, sensor dropout, or device removal differs by disease status, subgroup, or treatment arm | Patterns of data availability or modeling choices may distort estimates of the underlying physiology or treatment effect | Examine missingness and adherence by subgroup and study arm; use appropriate missing data and measurement error methods; report sensitivity analyses |
| Analysis or decision rule | Instability of multimodal inputs or data fusion procedures | A risk score changes when a wearable stream, patient report, imaging measure, or electronic health record input is unavailable | The conclusion may depend on site workflow or input availability rather than the intended clinical construct | Evaluate performance under missing or altered inputs; document data fusion procedures; assess calibration and stability across sites and workflows |
| Conclusion or action | Application to a different device, population, care setting, or workflow | Applying a measure validated in one population or clinical workflow to another | The evidence may provide weaker support when applied to a different population, deployment context, threshold, or clinical decision | Conduct external validation across devices, populations, settings, and workflows; evaluate subgroup performance; recalibrate thresholds when justified |
| Across the pathway | Firmware updates, software revisions, algorithm retraining, device replacement, or workflow changes over time | A measurement or data processing system changes after the original validation study | The measurement-to-conclusion pathway may drift, weakening support for the original scientific, clinical, or regulatory conclusion | Maintain version histories; monitor performance over time; define criteria for re-evaluation; repeat validation after changes likely to affect downstream conclusions |

## Applying the framework

The level of evaluation should match the consequence of the conclusion. Descriptive uses may need transparent reporting and simple sensitivity checks. A measure used to trigger an alert, define treatment response, select a study subgroup, support a pivotal trial conclusion, or support a product label needs stronger evidence because false positives, false negatives, and biased estimates have greater consequences. At minimum, researchers should state the measure, the conclusion attached to it, the measurement system features that could affect that conclusion, and how they evaluated uncertainty and subgroup performance.

Clinicians increasingly use CGM and other device-derived summaries alongside laboratory measures, symptoms, and treatment history when considering changes to a treatment regimen. Yet the reports available in practice may provide limited information about device version, systematic bias, software or algorithm updates, missing data, wear patterns, performance near decision thresholds, or subgroup-specific measurement behavior. Clinical reports and decision support systems could flag measurement system features most likely to affect interpretation, reducing the need for clinicians to track device and software changes across reports. A priority for future research is to determine when uncertainty is large enough to change clinical interpretation and how to communicate that information without adding unnecessary burden to clinical workflows.

**AI-generated and multimodal digital measures**

Inferential validity becomes especially important as digital health moves from single sensor summaries toward AI-generated and multimodal measures. Modern systems may combine wearable streams, electronic health records, patient reports, imaging, speech, environmental signals, and/or multiomics data into a risk score, digital biomarker, or digital twin representation[33]. These measures may depend on feature extraction, data fusion, procedures for handling missing data, retraining, device versioning, and software version changes rather than a single transparent physiologic summary[34-38].

High predictive accuracy can conceal weaknesses in the conclusions supported by a model when its performance depends on a particular device type, deployment setting, clinical workflow, pattern of care access, adherence behavior, or population composition. A model that performs well in one deployment context may not support the same conclusions when applied to different devices, populations, workflows, or healthcare settings. For AI-supported care and digital biomarker development, it is therefore important to report predictive accuracy, measurement system dependencies, deployment context (including devices, populations, care settings, and clinical workflows), subgroup performance, performance across deployment settings, sensitivity to preprocessing, and the effect of retraining or changes to measure construction on downstream conclusions.

**Design, reporting, and methodological implications**

Digital health technologies evolve. Devices and firmware are updated, sensors are replaced, operating systems change, missing data patterns shift, preprocessing protocols change, algorithms are revised and retrained, users change how they interact with devices, and technologies are applied in new settings or used for decisions with greater clinical, scientific, or regulatory consequences. Inferential validity therefore is a lifecycle property that requires ongoing evaluation. Validation evidence from one time point may no longer support the same conclusion after a firmware update, preprocessing change, new deployment population, shifted missingness pattern, or use in a decision with greater clinical, scientific, or regulatory consequences. Recording device and software versions, monitoring subgroup performance, documenting measurement system changes, and prespecifying when analyses should be repeated can make factors that may affect inferential uncertainty visible and manageable throughout the technology lifecycle. Several methodological traditions can support inferential validity. Measurement error models address bias when observed measures differ from target quantities[39,40]. Functional data methods can help describe high-frequency trajectories[24,25], although smoothing alone does not correct systematic measurement error. Longitudinal and missing data methods address repeated observations, incomplete data, differential adherence, and informative dropouts[41]. Causal inference methods clarify the scientific question of interest and the assumptions needed to connect a digital measure to a conclusion[42].

It is crucial to develop practical tools for making measurement system uncertainty visible before using digital measures for high consequence decisions. Priorities include monitoring measurement system drift over time, using sparse reference measures to check wearable outputs, carrying measurement uncertainty into threshold and treatment effect decisions, evaluating subgroup-specific measurement effects with limited validation data, and determining when updated measurement systems require re-evaluation.

## Conclusion

Digital measures are reshaping clinical research, remote monitoring, AI-supported care, and regulatory decision making. Their value depends on how well devices measure and how well the resulting measures support the conclusions that researchers and clinicians draw from them. Inferential validity makes this link explicit. A focus on inferential validity can help make digital medicine more trustworthy, interpretable, and useful by connecting measurement system evidence to treatment effect estimation, threshold decisions, subgroup interpretation, generalizability, and ongoing monitoring. As digital health technologies become increasingly central to evidence generation and clinical decision making, inferential validity provides a framework for ensuring that confidence in digital measures translates into confidence in the conclusions they support.

## Acknowledgments

This work was supported in part by the National Institute of Diabetes and Digestive and Kidney Diseases under Award Nos. R01DK132385 and R01DK136994. Support was also provided by the New York Regional Center for Diabetes Translation Research (P30DK111022). The content is solely the responsibility of the authors and does not necessarily represent the official views of the National Institutes of Health. The authors thank Drs. Lisa Giles and Devon Brewer for editorial assistance and for their detailed, constructive critique and feedback, which substantially improved the clarity, accessibility, and structure of the manuscript.

## Author contributions

C.D.T. and R.S.Z. conceived the concept of inferential validity and led manuscript development. C.D.T., M.O., Y.L., A.M., and R.S.Z. contributed to the conceptual framework, interpretation, and writing. O.M.B., S.L.L. and J.S.G. reviewed the manuscript for scientific and clinical relevance and interpretation. All authors reviewed, edited, and approved the final manuscript.

## Competing interests

Outside of this work, C.D.T. and R.S.Z. are co-founders of ZoTek Analytics, LLC, a company focused on advanced analytics for complex health data. Outside of the current work, OMB discloses that he received subcontract grants to Penn State from Mobile Sleep Technologies doing business as SleepSpace (NIH/NIA SBIR R43-AG056250, R44-AG056250), received honoraria/travel support for lectures from Tufts School of Dental Medicine, University of Utah, University of Arizona, Spencer Study Club, and University of Miami; consulting fees from Georgia State University, National Institute on Aging, and Harvard Chan School of Public Health; and received an honorarium from the National Sleep Foundation for his role as the Editor in Chief (2019-2025) of Sleep Health sleephealthjournal.org. The remaining authors declare no competing interests.

## Data availability

The illustrative analyses we presented were based on the publicly available National Health and Nutrition Examination Survey (NHANES)[20-23], CGMacros[27,28], and OpenOximetry datasets[29-31].

## Code availability

We will make code used for the illustrative analyses in this manuscript publicly available in a repository upon publication.